\documentclass[12pt]{iopart}
\usepackage{graphicx}
\usepackage{epstopdf}
\usepackage{subfigure}
\usepackage{epsfig}
\usepackage{amssymb,amsfonts}
\usepackage{color}
\begin{document}

\title[]{The Influence of Initial State Fluctuations on Heavy Quark Energy Loss in Relativistic Heavy-ion Collisions}

\author{Shanshan Cao $^1$, Yajing Huang $^{2}$, Guang-You Qin $^{3}$ and Steffen A Bass $^1$}
\address{$^1$ Department of Physics, Duke University, Durham, NC 27708, USA}
\address{$^2$ School of Physics, Shandong University, Jinan, 250100, China}
\address{$^3$ Institute of Particle Physics and Key Laboratory of Quark and Lepton Physics (MOE), Central China Normal University, Wuhan, 430079, China}
\ead{\mailto{bass@phy.duke.edu}}

\begin{abstract}

We study the effects of initial state fluctuations on the dynamical evolution of heavy quarks inside a quark-gluon plasma created in relativistic heavy-ion collisions. The evolution of heavy quarks in QGP matter is described utilizing a modified Langevin equation that incorporates the contributions from both collisional and radiative energy loss. The spacetime evolution of the fireball medium is simulated with a (2+1)-dimensional viscous hydrodynamic model. We find that when the medium traversed by the heavy quark contains a fixed amount of energy, heavy quarks tend to lose more energy for greater fluctuations of the medium density. This may result in a larger suppression of heavy flavor observables in a fluctuating QGP matter than in a smooth one. The possibility of using hard probes to infer the information of initial states of heavy-ion collisions is discussed.

\end{abstract}

\maketitle

%\date{\today}

%%%%%%%%%%%%%%%%%%%%%%%%%%%%%%%%%%%%%%%%%%%%%%%%%%%%%%%%%%%%%%%%%%%%%

\section{Introduction}
\label{sec:Introduction}

High energy heavy-ion collisions at the Relativistic Heavy-ion Collider (RHIC) and the Large Hadron Collider (LHC) provide unique opportunities to study QCD matter at extreme temperatures and densities. It is now well established that a highly excited and color deconfined quark-gluon plasma (QGP) may be created in these energetic nucleus-nucleus collisions. The dynamical evolution of the bulk matter has been successfully described by models utilizing relativistic hydrodynamic simulation \cite{Kolb:2003dz}, in particular the strong collective flow behaviors exhibited by the soft hadrons formed out of the expanding QGP fireball \cite{Adams:2003zg,Aamodt:2010pa,ATLAS:2011ah}.

Recently there has been significant effort in studying initial state fluctuations in heavy-ion collisions, such as the fluctuations of nucleon positions and color charges inside initial colliding nuclei \cite{Luzum:2013yya}.
Some of the most interesting consequences of initial state fluctuations include nonzero anisotropy in ultra-central collisions and the presence of odd-order harmonic geometry and flow \cite{Alver:2010gr, Petersen:2010cw, Staig:2010pn, Qin:2010pf, Ma:2010dv, Xu:2010du, Teaney:2010vd, Qiu:2011iv, Bhalerao:2011yg, Floerchinger:2011qf}.
The elliptic, triangular and other higher-order harmonic flows have been measured at RHIC and the LHC \cite{Adare:2011tg, Adamczyk:2013waa, ATLAS:2012at}.
These measurements have triggered great interest in studying the origin of the initial state fluctuations, and how they affect the dynamical evolution of the fireball and manifest in the final state particle flow and correlations \cite{Petersen:2010cw, Qin:2010pf, Teaney:2010vd, Qiu:2011iv, Qin:2011uw, Qiu:2012uy, Pang:2012uw, Gale:2012rq}.
One of the purposes of these studies is to obtain a quantitative extraction of transport properties such as the shear viscosity of the QGP matter produced in high energy nucleus-nucleus collisions.

Interestingly, the initial conditions especially the geometry of the heavy-ion collisions still remain one of the major uncertainties in the extraction of a precise value for QGP shear viscosity \cite{Luzum:2008cw, Dusling:2007gi, Song:2010mg}.
Various fluctuations such as the initial transverse flow and longitudinal fluctuations \cite{Pang:2012he, Pang:2012uw}, and the medium response to the jet energy loss \cite{Andrade:2014swa} may introduce more uncertainties in our understanding of the initial states. The purpose of this work is to investigate the effect of fluctuating initial conditions on the dynamics of heavy quark in medium and whether it is possible to infer information on the initial state fluctuations in heavy-ion collisions from heavy flavor observables.

Heavy quarks, due to their large masses, are mainly produced via initial hard scatterings, and thus provide a valuable tool to probe the spacetime profile and transport properties of the QGP fireball.
Previous studies have shown that low-$p_\mathrm{T}$ heavy quarks provide direct measure of the thermal properties of the medium, while at large $p_\mathrm{T}$ heavy flavor quarks may provide a reference to investigate the medium modification of high-energy jets \cite{vanHees:2005wb,Rapp:2009my}.
At intermediate $p_\mathrm{T}$, heavy quarks and mesons may provide rich information for our understanding of fragmentation-versus-coalescence mechanisms for hadron formation \cite{Lin:2003jy,Greco:2003vf,Oh:2009zj,He:2011qa,Cao:2013ita}.
From the experimental side, RHIC and LHC have observed significant suppression at high $p_\mathrm{T}$ and strong elliptic flow for heavy flavor mesons and heavy flavor decay electrons \cite{Adare:2010de,Tlusty:2012ix,Grelli:2012yv,Caffarri:2012wz}.
Various transport models have been developed to study the in-medium evolution of heavy quarks, such as the Boltzmann-based parton cascade model (BAMPS) \cite{Uphoff:2012gb}, the linearized Boltzmann model coupled to a hydrodynamic medium \cite{Gossiaux:2010yx,Nahrgang:2013saa}, and the Langevin evolution of heavy quark inside the QGP medium \cite{Moore:2004tg, Akamatsu:2008ge,He:2011qa,Young:2011ug,Alberico:2011zy,Lang:2012cx,Cao:2011et,Cao:2012jt,Cao:2013ita}.

In studying heavy quark evolution and energy loss in realistic hydrodynamic QGP matter, smooth initial conditions have been utilized for hydrodynamical evolution in most literatures.
The influence of the initial state fluctuations on heavy quarks has not yet been studied.
There have been similar studies of the effect of the initial state fluctuations in the context of high $p_\mathrm{T}$ light flavor jets \cite{Rodriguez:2010di,Renk:2011qi,Zhang:2012ik}, but no unified conclusion has been reached yet. For instance, Ref. \cite{Rodriguez:2010di} used a (1+1)-dimensional Bjorken hydrodynamic background and found that the fluctuation in the spatial distribution of the initial hard scatterings reduces the suppression of jet production in the limit of strong, surface-dominated quenching. In Ref. \cite{Zhang:2012ik} it was found that with the inclusion of the transverse expansion of the medium, i.e., using a (2+1)-dimensional hydrodynamic model, jet energy loss is enhanced when the initial state fluctuation is incorporated. A detailed investigation has also been implemented in Ref. \cite{Renk:2011qi}, where the angular dependence of parton suppression, the correlation between parton production points and hot spot locations, and effects of different energy loss models have been discussed. Using a (2+1)-dimensional medium, they found that due to the cancellation between various effects, parton suppression displays no significant difference in central collisions between fluctuating and smooth initial conditions, but is slightly reduced in peripheral collisions when the medium is fluctuating.

In this work, we investigate the influence of the initial state fluctuations on heavy quark evolution inside the QGP matter. We simulate the dynamical evolution of heavy quarks using our modified Langevin equation developed in Ref. \cite{Cao:2013ita} that includes both collisional and radiative energy loss mechanisms.
The QGP medium is simulated with a (2+1)-dimensional viscous hydrodynamic model which has been tuned to describe the bulk matter observables.  To isolate the effects of the initial state fluctuations on heavy quark energy loss, we do not include the hadronization process and the subsequent interaction between heavy mesons and hadron gas in this work. We do not aim for a direct comparison with experimental data in this work, but focus on exploring how heavy quark evolution and energy loss are affected by the the presence of the initial state fluctuations. One may refer to Ref. \cite{Cao:2013ita} for a detailed comparison to experimental data using smooth initial conditions in which the hadronization process of heavy quarks is included. We will discuss the possibility of utilizing heavy quarks as an additional tool to provide the information about the granularity of local fluctuations inside QGP, apart from those traditional observables such as higher-order harmonic flow coefficients of soft hadrons.

The paper is organized as follows. In Sec.\ref{sec:HQEvolution}, following our previous work \cite{Cao:2013ita}, we briefly introduce our Langevin approach for simulating the dynamical evolution of heavy quarks inside QGP. In Sec.\ref{sec:EnergyLoss}, we study how the energy loss of heavy quark is affected by the size and the number of local fluctuations (hot spots) using a static medium. In Sec.\ref{sec:QuenchingFlow}, we study the influence of the initial state fluctuations on heavy quark quenching using a realistic hydrodynamic medium. The summary and discussion will be presented in Sec.\ref{sec:summary}.

\section{Heavy Quark Evolution inside QGP}
\label{sec:HQEvolution}

In the limit of multiple scatterings, the in-medium evolution of heavy quarks can be treated as Brownian motion and is usually described using the Langevin approach.
Apart from the collisional energy loss resulting from quasi-elastic scatterings, heavy quarks may also lose energy through medium-induced gluon radiation.
To include both contributions, we use the modified Langevin equation developed in our previous work \cite{Cao:2013ita}:
\begin{equation}
\label{eq:modifiedLangevin}
\frac{d\vec{p}}{dt}=-\eta_D(p)\vec{p}+\vec{\xi}+\vec{f}_g.
\end{equation}
The first two terms on the right-hand side represent the drag force and the thermal random force. The last term $\vec{f}_g=-d\vec{p}_g/dt$ is introduced to describe the recoil force exerted on heavy quarks due to gluon radiation, where $\vec{p}_g$ denotes the momentum of radiated gluons.

In our simulation, we determine the probability of gluon radiation during a time interval $\Delta t$ using the average number of radiated gluons:
\begin{equation}
\label{eq:gluonnumber}
P_\mathrm{\scriptsize{rad}}(t,\Delta t)=\langle N_g(t,\Delta t)\rangle = \Delta t \int dxdk_\mathrm{T}^2 \frac{dN_g}{dx dk_\mathrm{T}^2 dt}.
\end{equation}
We choose sufficiently small time steps $\Delta t$ to ensure that this probability is smaller than one in a single evolution time step.
Here the medium-induced gluon emission spectrum is taken from the higher-twist energy loss formalism \cite{Guo:2000nz,Majumder:2009ge,Zhang:2003wk}:
\begin{eqnarray}
\label{eq:gluondistribution}
\frac{dN_\mathrm{g}}{dx dk_\mathrm{T}^2 dt}=\frac{2\alpha_s  P(x)\hat{q} }{\pi k_\mathrm{T}^4} {\sin}^2\left(\frac{t-t_i}{2\tau_f}\right)\left(\frac{k_\mathrm{T}^2}{k_\mathrm{T}^2+x^2 M^2}\right)^4,
\end{eqnarray}
where $k_\mathrm{T}$ is the transverse momentum of the radiated gluon, and $x$ is the fractional energy taken away by the radiated gluon. Additionally, $\alpha_s$ is the strong coupling constant, $P(x)$ is the parton splitting function and $\hat{q}$ is the gluon transport coefficient defined as transverse momentum broadening per unit length -- $\hat{q}\equiv d\langle\mathrm{\Delta}p_\mathrm{T}^2\rangle/dt$.
The gluon formation time is defined as $\tau_f={2Ex(1-x)}/{(k_\mathrm{T}^2+x^2M^2)}$, in which $E$ and $M$ are the energy and mass of heavy quarks.
At a given time step, Eq. (\ref{eq:gluonnumber}) is utilized to determine the probability of gluon formation.
If a gluon is radiated, its energy and momentum is then sampled using the Monte-Carlo (MC) method according to the gluon spectrum given by Eq. (\ref{eq:gluondistribution}). Note that in this work, $p$ is used for heavy quark momentum and $k$ for radiated gluon momentum.

With the requirement that heavy quarks approach thermal equilibrium after sufficiently long time of in-medium evolution, the fluctuation-dissipation relation between the drag and the thermal force may be obtained -- $\eta_D(p)=\kappa/(2TE)$, where $\kappa$ is called the momentum space diffusion coefficient defined as $\langle\xi^i(t)\xi^j(t')\rangle=\kappa\delta^{ij}\delta(t-t')$. Note that for the radiative energy loss we have only included the gluon emission at this moment but not the inverse process. A lower cut-off for the radiated gluon energy $\omega_0=\pi T$ is implemented to provide a separation between the non-equilibrium regime of high-$p_T$ radiative energy-loss and the near-thermal QGP regime in which the balance between gluon emission and absorption would be important. Above $\omega_0$, the classical Langevin equation is modified by gluon radiation; but below the cut-off, the heavy quark motion is entirely governed by collisional energy loss and heavy quarks are able to thermalize in the medium. In this work, we relate different transport coefficients via $D=2T^2/\kappa$ and $\hat{q}=2\kappa C_A/C_F$, where $D$ is the spatial diffusion coefficient of heavy quarks.
With the above setup, there is only one free parameter in our Langevin framework.
Throughout our calculation below, the spatial diffusion coefficient of heavy quarks is chosen as $D=6/(2\pi T)$, which is equivalent to $\hat{q}$ around 3~GeV$^2$/fm at a temperature of 400~MeV.
Such setup has been shown to provide the best $p_\mathrm{T}$ spectra of $D$ meson $R_\mathrm{AA}$ in 2.76~TeV central Pb-Pb collisions \cite{Cao:2013ita}.
We note that the value of jet transport coefficient used here is comparable to many other studies on light flavor jet quenching \cite{Burke:2013yra, Qin:2010mn}.

Using our modified Langevin framework, we may simulate the heavy quark evolution inside a QGP matter produced in relativistic heavy-ion collisions.
For the space-time evolution of the QGP fireballs, a (2+1)-dimensional viscous hydrodynamic model is used \cite{Song:2007fn,Song:2007ux,Qiu:2011hf}. Here we employ the code version and parameter tunings taken from Ref. \cite{Qiu:2011hf}.
A MC-Glauber model is adopted to generate the positions of participant nucleons and binary collisions, providing both the initial conditions of hydrodynamics and the spatial distribution of initial heavy quarks.
The initial momentum spectra of heavy quarks is obtained from a leading-order perturbative QCD calculation.
The hydrodynamic model provides the spacetime evolution of the local temperature and collective flow of the QGP medium. For every Langevin time step, we first boost the heavy quark into the rest frame of the fluid cell, and the energy and momentum of the heavy quark are updated according to the above Langevin algorithm before it is boosted back to the global center-of-mass frame.

In our simulation, heavy quarks are assumed to stream freely prior to hydrodynamics' starting time $\tau_0=0.6$~fm/c. The possible energy loss during the pre-equilibrium stage is neglected; it is expected to be small due to its short period of time compared to the much longer history of the fireball evolution.
Once they enter a fluid cell whose local temperature is below 165~MeV, heavy quarks are treated as free-streaming again.

\section{Effects of Fluctuations on Heavy Quarks: A Static Medium Case}
\label{sec:EnergyLoss}

In this section, we investigate the influence of the local temperature fluctuations (or hot spots) on heavy quark energy loss in a static QGP medium. We will look at two different aspects of density fluctuations: the size and the number of local fluctuations. To mimic the effect of the realistic (2+1)-dimensional boost invariant hydrodynamic medium which we will use in the next section, the static medium is chosen to be two dimensional, i.e., the hot spots are in fact hot tubes in this case.

\begin{figure}[tb]
  \centering
  \epsfig{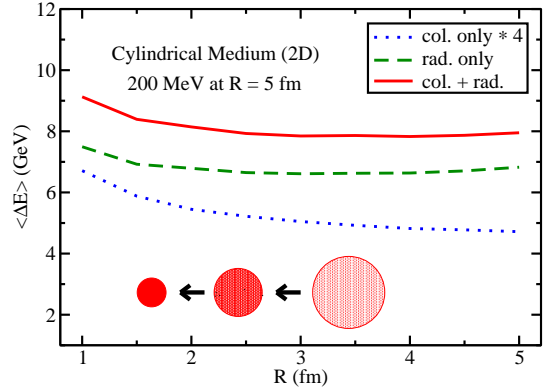}
  \caption{(Color online) Energy loss of charm quark as a function of the size of the hot tube.}
 \label{fig:hotspot_2Dsize}
\end{figure}

For the first scenario, we generate one cylindrical medium (hot tube) with a constant temperature. As demonstrated by the cartoon inside Fig. \ref{fig:hotspot_2Dsize}, we vary its size and study how the energy loss of charm quarks is affected. When varying the size, the total energy contained inside the tube is kept fixed.
The temperature of the medium is set as 200~MeV when the tube radius is $R=5$~fm and will increase as the radius decreases. Each charm quark is initialized with 50~GeV and placed at the center of the cylinder.
We calculate the average energy loss of charm quarks as they exit the hot tube medium, and the results are shown in Fig. \ref{fig:hotspot_2Dsize}. We also compare the results using different energy loss mechanisms of heavy quarks: collisional energy loss only, radiative only and the combined loss.
In the plot we multiply the results from pure collisional energy loss by a factor of 4 for a better resolution.
From the figure, we observe that the energy loss of charm quarks is not very sensitive to the size of the hot tube (with the total medium energy unchanged).

\begin{figure}[tb]
  \centering
  \epsfig{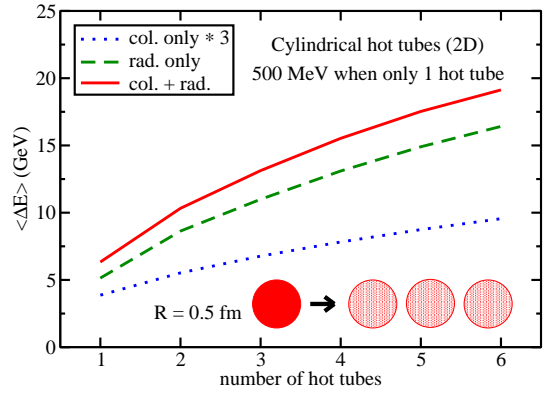}
  \caption{(Color online) Energy loss of charm quark as a function of the number of hot tubes.}
 \label{fig:hotspot_2Dnum}
\end{figure}

To study the effect of the number of local density fluctuations on the heavy quark energy loss, we generate $N$ hot tubes with the same radius $R=0.5$~fm. As displayed by the cartoon inside Fig. \ref{fig:hotspot_2Dnum}, they are lined up along charm quarks' initial propagation direction. The initial charm quark energy is set as 50~GeV (placed at the edge of the first hot tube) and the temperature of the medium is set as 500~MeV when there is only one hot tube. Again, when changing the number of hot tubes, the total energy contained in the medium (sum of the $N$ hot tubes) is fixed. The result for such scenario is shown in Fig. \ref{fig:hotspot_2Dnum}. We see that the energy loss of charm quarks is quite sensitive to the number of hot tubes. Note that similar to Fig. \ref{fig:hotspot_2Dsize}, an additional factor (3 here) is multiplied to the result of pure collisional energy loss just for a better resolution.

The above results can be easily understood with the following argument.
One may assume the power law dependence for heavy quark energy loss on the medium length and temperature as follows:
\begin{equation}
 \label{eq:EnergyLoss1}
 \Delta E \propto (NR)^\alpha T^\beta,\hspace{9pt} \epsilon\propto T^4,\hspace{9pt} V\propto NR^d,
\end{equation}
where $N$ is the number of hot tubes, $R$ is the radius of each hot tube probed by heavy quarks, and $V$ is the total volume of the $d$-dimensional medium. $T$ is the temperature, and $\epsilon$ is the medium energy density.
The parameters $\alpha$ and $\beta$ denote power law dependence of heavy quarks on the path length and the medium temperature.
Since we fix the total amount of energy contained in the medium, i.e., $\epsilon V=\mathrm{Const.}$, one may obtain the dependence on the size and the number of hot tubes as:
\begin{equation}
 \label{eq:EnergyLoss2}
 \Delta E \propto N^{\alpha-\beta/4}R^{\alpha-\beta d/4}.
\end{equation}
In our energy loss model, with the assumption that the diffusion coefficient $D$ is inversely proportional to $T$, we extract from Monte-Carlo simulations of heavy quark propagation in static medium that $\alpha=1$ and $\beta=2$ for collisional energy loss and $1<\alpha<2$ (say taking $\alpha=3/2$ in the following analysis) and $\beta=3$ for radiative energy loss.

When there is only one hot tube $N=1$ (the first scenario), Eq. (\ref{eq:EnergyLoss2}) is reduced to $\Delta E \propto R^{\alpha-\beta d/4}$. Thus with the values of $\alpha$ and $\beta$ above, $\Delta E$ should be in principle independent of $R$ in a 2-dimensional system for both collisional and radiative energy loss. This can be understood as a balance between the two effects of varying the system size of a 2-dimensional system: by decreasing $R$, heavy quarks propagate through a shorter path length which decreases their energy loss; on the other hand, the energy density increases which enhances their energy loss. In Fig. \ref{fig:hotspot_2Dsize}, the simulation results indicate that heavy quark energy loss is indeed insensitive to the system size although it is not strictly a constant because the above extracted values of $\alpha$ and $\beta$ are only an overall estimation but not exact. For a 1-dimensional system, we have verified that the total energy loss of heavy quarks decreases when confining the same amount energy in a smaller region because the increase of the energy density is slower and its effect is overwhelmed by shortening the path length. To the contrary, the energy loss of heavy quarks increases in a 3-dimensional system when the system size is decreased.

Similarly, one may fix the the size $R$ of hot tubes in Eq. (\ref{eq:EnergyLoss2}) to isolate the influence of the number of hot tubes: $\Delta E \propto N^{\alpha-\beta/4}$.
One can see that the total energy loss does not depend on the dimension of the system, but increases significantly with the number of hot tubes for both collisional and radiative energy loss.
This is consistent with the finding shown in Fig. \ref{fig:hotspot_2Dnum}.
Note that Fig. \ref{fig:hotspot_2Dsize} and Fig. \ref{fig:hotspot_2Dnum} are designed to study two different aspects of the initial-state fluctuation and are not expected to compared between each other. Moreover, there is no specific choice of the initial radius $R$ and temperature $T$ in our toy-model studies,  except that when $R$ and $N$ vary, $T$ is expected to be kept in the range of the realistic QGP temperature reached at the RHIC and LHC experiments.

\begin{figure}[tb]
  \centering
  \epsfig{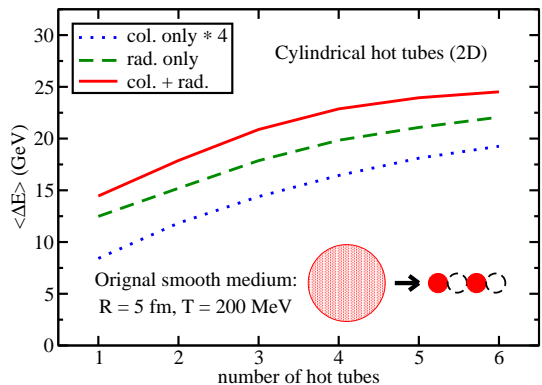}
  \caption{(Color online) Effects of the strength of medium fluctuation (number of hot tubes $N$) on charm quark energy loss.}
 \label{fig:hotspot_2Dcombine}
\end{figure}

One may combine the above two scenarios, i.e., changing the size and number of hot spots/tubes simultaneously. This is very similar to the change from a large and smooth medium to fluctuating medium consisting of several hot (and cold) regions as demonstrated by the cartoon inside Fig. \ref{fig:hotspot_2Dcombine}. The total energy contained in these two different media are the same.
To simplify the study, we split a large smooth tube medium into $N$ hot tubes with smaller sizes, which are lined up adjacent to each other along the direction of the initial momentum of our charm quarks ($E_{\mathrm{init}}=50$~GeV). Another $N$ cold tubes (vacuum here) are also placed between every two hot tubes to mimic the realistic distribution of local density fluctuations. The sizes of the smaller tubes are chosen such that the total length $4NR$ traversed by heavy quarks is fixed as the diameter of the original smooth medium with a radius of 5~fm and temperature of 200~MeV. Note that heavy quarks stream freely in the cold tubes here.
The results for a 2-dimensional system are shown in Fig. \ref{fig:hotspot_2Dcombine}.
One observes that the energy loss of charm quarks increases when the original smooth medium is split into more hot and cold tubes, i.e., the more fluctuations the medium has, the stronger energy loss the charm quarks experience.

\begin{figure}[tb]
  \centering
  \epsfig{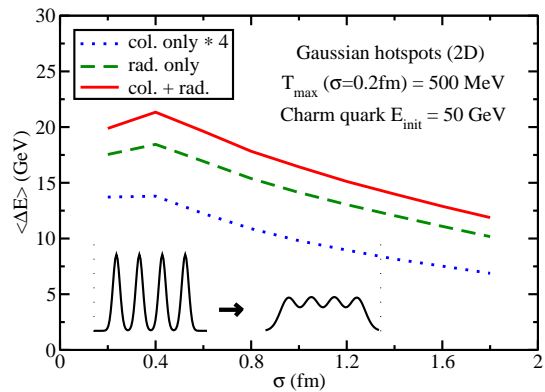}
  \caption{(Color online) Effects of the strength of medium fluctuation (width $\sigma$ of each Gaussian shape hot tube) on charm quark energy loss.}
 \label{fig:hotspot_2D_gauss}
\end{figure}

In Fig. \ref{fig:hotspot_2D_gauss}, we investigate the influence of medium fluctuation on charm quark energy loss with a more realistic picture. Four hot tubes are lined along the $x$ axis with their centers fixed at 2, 4, 6  and 8~fm. The energy density distribution of each hot tube obeys a 2-dimensional Gaussian distribution in the $x-y$ plane. We then vary the widths of these Gaussian hot tubes while keeping the total medium energy constant, and study the influence on the total energy loss of the heavy quark. The normalization of each Gaussian distribution is determined such that its central temperature is 500~MeV when the width is $\sigma=0.2$~fm, and heavy quarks are initialized at $x=0$ with 50~GeV energy along the $x$ direction. Note that when $\sigma$ is small (less than 0.8~fm), four peaks of hot tubes are separated from each other. However, they begin to overlap when $\sigma$ becomes large, which indicates that the medium is significantly smoothed. We observe from Fig. \ref{fig:hotspot_2D_gauss} that the total energy loss of charm quark decreases with increasing Gaussian width $\sigma$, suggesting that the charm quark tends to lose more energy when the medium density along its propagation path has stronger fluctuations. The slightly non-monotonic behavior in Fig. \ref{fig:hotspot_2D_gauss} at small $\sigma$ is due to the transverse momentum broadening experienced by the heavy quark which makes its path deviate from the $x$-axis. With the chosen transport coefficient, a heavy quark is able to gain transverse momentum over 1~GeV after traversing the first two hot tubes and thus miss the dense regions of the last few hot tubes if $\sigma$ is sufficiently small.

To sum up for this section, we find that the energy loss of charm quarks in a 2-dimensional system does not have much dependence on the size of the local fluctuations, but is quite sensitive to the number of local fluctuations in the medium. When the total amount of energy contained in the medium is fixed along their path, heavy quarks tend to lose more energy in a fluctuating medium than in a smooth one. Although the above results are obtained using a static medium, it provides an intuition to explain the results for a realistic hydrodynamic medium presented in the next section. We also note that our finding is based on the path length and temperature dependence of heavy quark energy loss in our model, i.e., the values of $\alpha$ and $\beta$ in Eq. (\ref{eq:EnergyLoss2}). In addition, the above conclusions depend on the assumption made in our toy model that the total energy deposited by heavy-ion collisions is fixed while the volume in which it is deposited fluctuates; other reasonable modelings of the initial state fluctuations, such as fixing the average local energy density, entropy density or temperature, could lead to different results. Also in our toy model, we initialize each given heavy quark inside the hot region and arrange the hot tubes along its initial direction. If these correlations between initial heavy quarks and the hot tubes are not satisfied, the energy loss could be reduced as discussed in Ref. \cite{Renk:2011qi}.

\section{Heavy Quarks in Event-by-Event Hydrodynamic Medium}
\label{sec:QuenchingFlow}

\begin{figure}[tb]
 \centering
 \subfigure{\label{fig:plotEvent}
      \epsfig{file=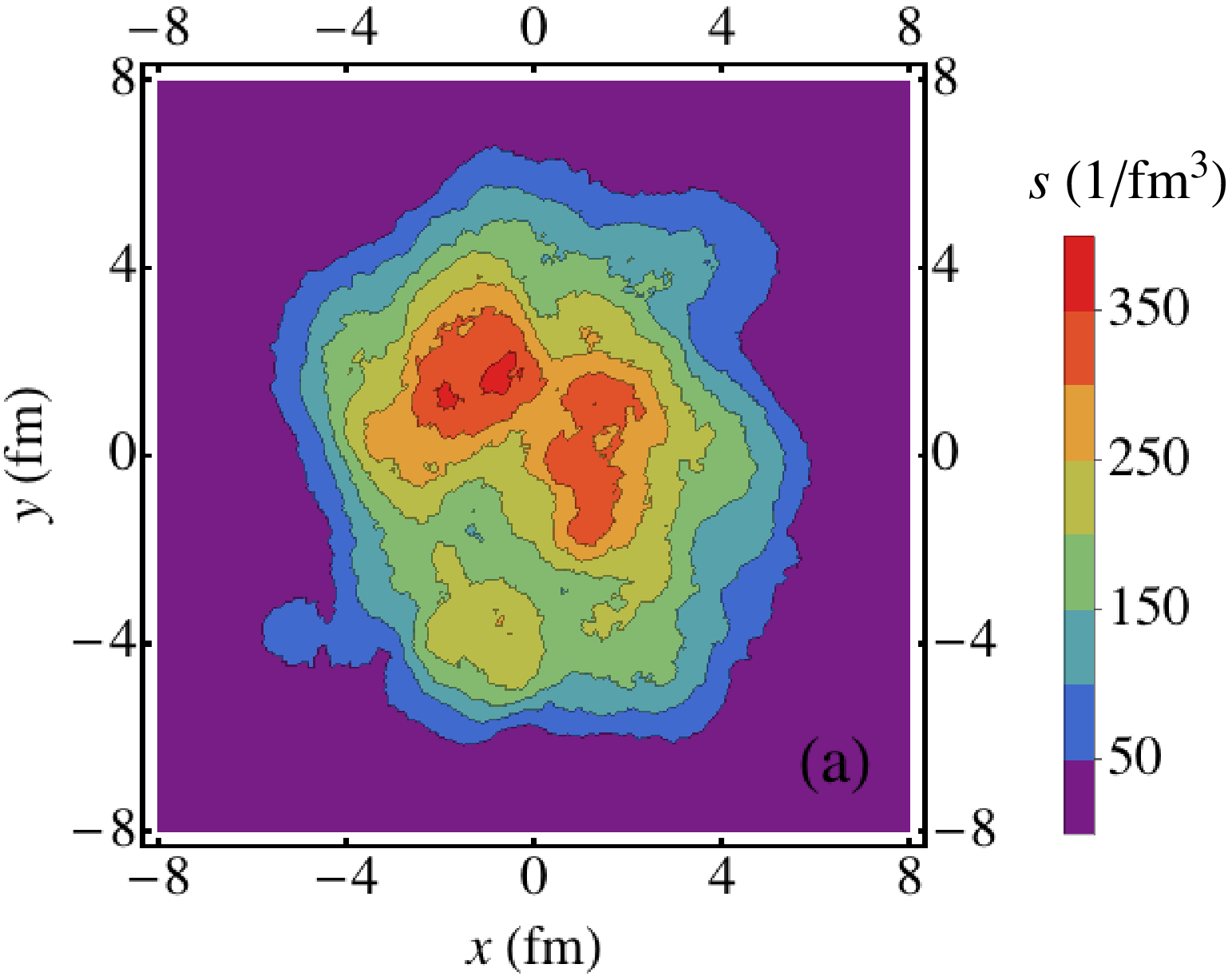, width=0.48\textwidth, clip=}}
 \subfigure{\label{fig:plotSmooth}
      \epsfig{file=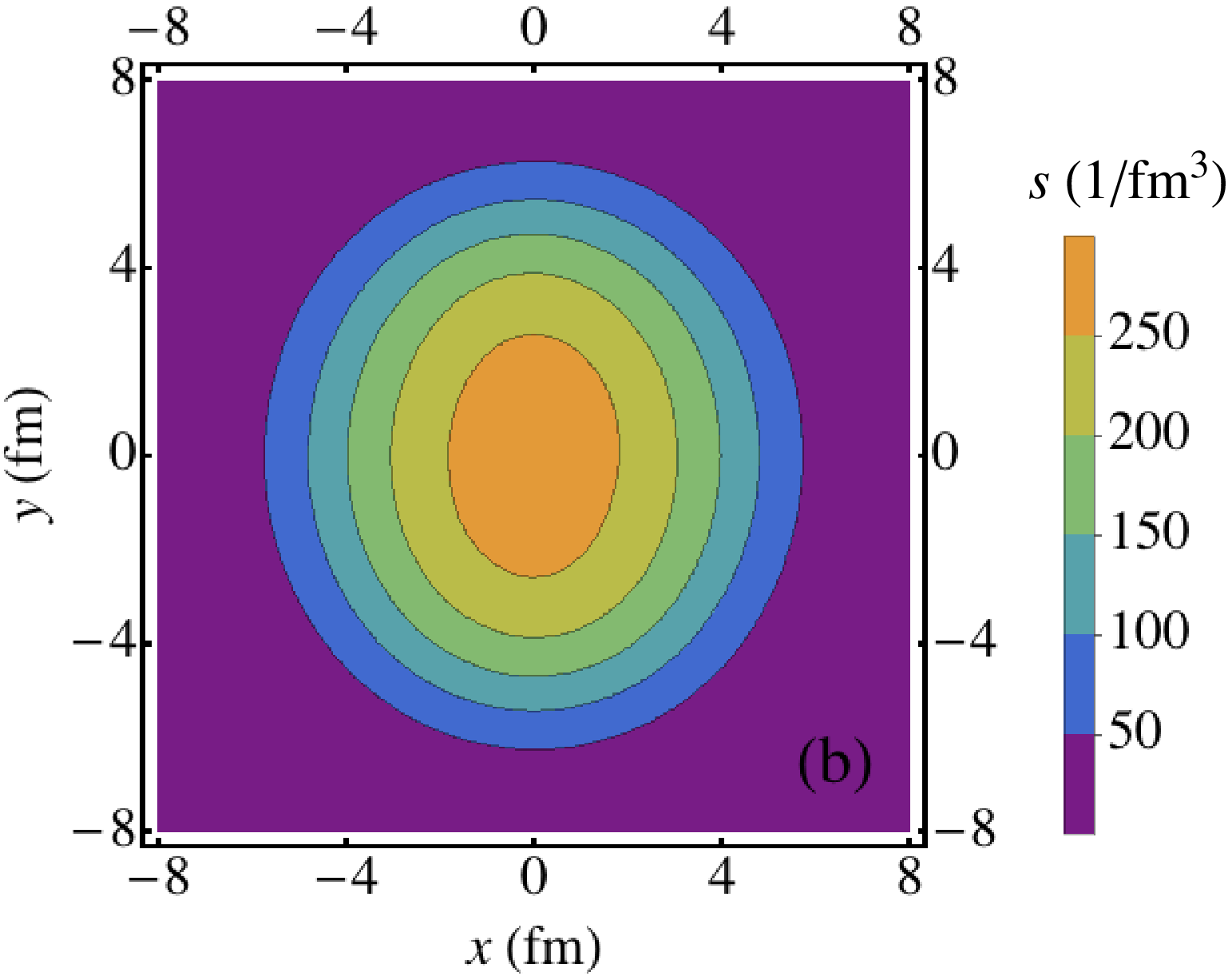, width=0.48\textwidth, clip=}}
 \caption{(Color online) Comparison between (a) a typical fluctuating event and (b) a smooth initial entropy density profile of hydrodynamical evolution of 2.76~TeV central Pb-Pb collisions.}
 \label{fig:InitialProfiles}
 \end{figure}

In the previous section, we have studied the response of heavy quark energy loss to the temperature fluctuations in a static QGP medium. In this section, we perform the investigation for a realistic expanding medium in which both temperature fluctuations and flow (fluctuations) are present.
Here, we utilize a (2+1)-dimensional viscous hydrodynamic model as described in Sec.\ref{sec:HQEvolution} to simulate the dynamical evolution of the hot QGP produced in Pb-Pb collisions at the LHC.
The initial conditions for the hydrodynamical evolution are obtained from the Monte-Carlo Glauber model.

In Fig. \ref{fig:InitialProfiles}, we compare the initial entropy density distribution in the transverse plane from a typical fluctuating event [Fig. \ref{fig:plotEvent}] and a smooth one generated by averaging over 100,000 fluctuating initial profiles for 0-7.5\% Pb-Pb collisions at 2.76 TeV at the LHC [Fig. \ref{fig:plotSmooth}]. We note that in Fig. \ref{fig:plotSmooth}, the initial profiles of all the events have been rotated to the same second-order participant plane before performing the event average of the entropy density. One can clearly see the presence of hot and cold regions in the QGP fireball for fluctuating initial conditions.

\begin{figure}[tb!]
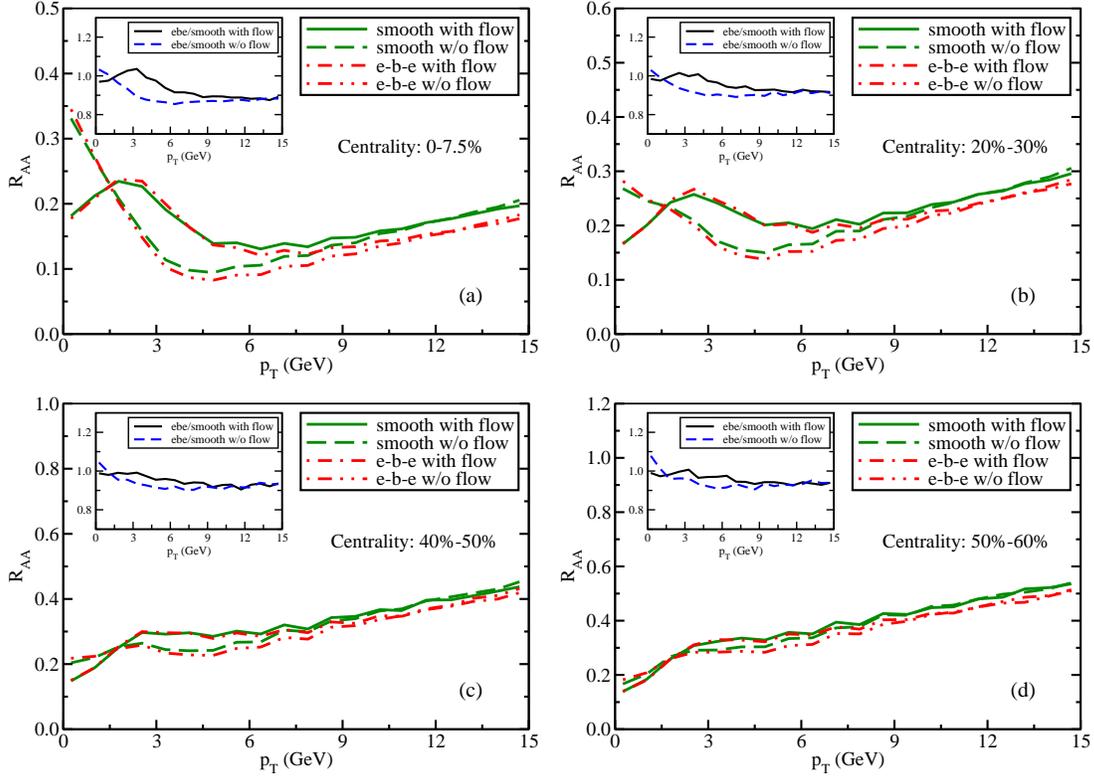

 \centering
 \subfigure{\label{fig:RAA-0-7d5}
      \epsfig{file=RAA-0-7d5.eps, width=0.45\textwidth, clip=}}
 \subfigure{\label{fig:RAA-20-30}
      \epsfig{file=RAA-20-30.eps, width=0.45\textwidth, clip=}}
 \subfigure{\label{fig:RAA-40-50}
      \epsfig{file=RAA-40-50.eps, width=0.45\textwidth, clip=}}
  \subfigure{\label{fig:RAA-50-60}
      \epsfig{file=RAA-50-60.eps, width=0.45\textwidth, clip=}}
 \caption{(Color online) Comparison of charm quark $R_\mathrm{AA}$ between calculations with smooth and fluctuating initial conditions of hydrodynamical evolution.}
 \label{fig:suppression}
\end{figure}

In Fig. \ref{fig:suppression}, we show the nuclear modification factor $R_\mathrm{AA}$ of charm quarks after their traveling through the QGP medium created in 2.76~TeV Pb-Pb collisions.
We compare the results from smooth initial conditions and from an event-by-event calculation for four different centralities. 
One can read from Fig. \ref{fig:RAA-0-7d5} - \ref{fig:RAA-50-60} that at high $p_\mathrm{T}$ the event-by-event calculations give larger suppression for heavy quarks than calculations with smooth initial conditions, i.e., the initial state fluctuations lead to larger energy loss for heavy quarks. As a consequence, slightly smaller suppression is observed for low $p_\mathrm{T}$ charm quarks with fluctuating initial conditions. One reason for such result may be understood from the discussion presented in Sec.\ref{sec:EnergyLoss}, i.e., the stronger fluctuations along the propagation path of heavy quarks tend to increase the energy loss when the total energy of the bulk matter is fixed (e.g., for a given centrality).

As has been mentioned, there exist temperature fluctuations and flow (fluctuations) in a realistic medium.
To remove and investigate the effect of the medium flow on heavy quark evolution, one may solve the Langevin equation Eq. (\ref{eq:modifiedLangevin}) in the global center-of-mass frame instead of the local rest frame of the fluid cell \cite{Cao:2012jt}. In this way, the evolution of heavy quarks is solely affected by the temperature distribution and fluctuations of the medium. One can see that the effect of the medium flow is to boost low $p_\mathrm{T}$ charm quarks into medium $p_\mathrm{T}$ regime and form the bump structure for the nuclear modification factor $R_\mathrm{AA}$. This bump feature disappears when the flow is turned off in the calculation.

The above observation can be seen more clearly in the subfigures inside Fig. \ref{fig:RAA-0-7d5} - \ref{fig:RAA-50-60} where we show the ratios between the final state $p_\mathrm{T}$ spectra of charm quarks from the event-by-event calculations and those from the smooth cases. For the central collisions [Fig. \ref{fig:RAA-0-7d5}], we obtain about 12\% more quenching at high $p_\mathrm{T}$ for the fluctuating initial condition as compared to the smooth initial condition. This could result in a 10\%-15\% difference in the extraction of the gluon transport coefficient $\hat{q}$ inside QGP. For more peripheral collisions, the effect of initial state fluctuations on heavy quark energy loss is less; the quenching increases about 7\% when switching from the smooth to the fluctuating initial condition in 50\%-60\% Pb-Pb collisions [Fig. \ref{fig:RAA-50-60}].
It is worth noticing that before drawing firm conclusions on the initial-state fluctuations from the heavy quark $R_\mathrm{AA}$ here, it is important to constrain our understanding of other uncertainties, such as the in-medium fragmentation function of heavy quarks, the correlation between heavy quark production vertices and hot spot positions \cite{Renk:2011qi}, the different length dependences of heavy quark energy loss suggested by different formalisms [see Eq. (\ref{eq:EnergyLoss2})], and the detailed properties of the hydrodynamical evolution \cite{Zhang:2012ik}. As shown in our subfigures, the difference between results calculated with fluctuating and smooth initial conditions is also affected by the QGP flow especially at low $p_\mathrm{T}$.

\section{Summary}
\label{sec:summary}

In this work, we have studied the impact of initial state fluctuations on heavy quark evolution and energy loss  in relativistic heavy-ion collisions.
The in-medium evolution of heavy quarks is described using a modified Langevin equation that simultaneously incorporates collisional and radiative energy loss components.
We have investigated the effect of local fluctuations for both static and realistic expanding QGP media.
For realistic medium, we have utilized a (2+1)-dimensional viscous hydrodynamic model to simulate the spacetime evolution of the QGP fireball.
The initial conditions for hydrodynamics are obtained from a MC-Glauber model, for both smooth and fluctuating cases.

We have studied the effects of temperature fluctuations on heavy quark energy loss in terms of the sizes and the number of local fluctuations (hot spots). We found that the total energy loss of heavy quarks is not very sensitive to the sizes of local fluctuations in a 2-dimensional system, but the energy loss increases significantly with the increasing number of hot spots. With a fixed amount of energy contained in the medium, heavy quark tends to lose more energy when the energy density along its trajectory has stronger fluctuations. This may result in a larger suppression of heavy flavor spectra. And our simulation in a realistic QGP medium have demonstrated that fluctuating initial conditions may bring around 10\% more suppression for inclusive charm quark production at high $p_\mathrm{T}$ in relativistic nucleus-nucleus collisions. The effect of initial state fluctuations on heavy quark energy loss tends to diminish for more peripheral collisions. It should be emphasized that in order to utilize heavy quark quenching as a reliable probe of the initial state, we need to improve the constraints on other sources of uncertainties. As has been discussed in Sec.\ref{sec:EnergyLoss} and \ref{sec:QuenchingFlow}, the assumptions about the initial state fluctuation may affect the final heavy quark energy loss, and how the initial phase space of heavy quarks are correlated with the hot spots created in heavy-ion collisions will also influence the realistic heavy flavor suppression.

Our study constitutes an important contribution to the quantitative understanding of heavy quark dynamics in relativistic heavy-ion collisions with initial state fluctuations. Although we utilize heavy quarks in our study to probe the effects of the fluctuations, many of our results should apply to light flavor partons as well. Our results suggest that jet modification might be utilized to probe the fluctuations of QGP medium, such as the degree of inhomogeneity or the number of hot spots. We further note that the sensitivity of heavy quark energy loss to hot spot number might be enhanced when one uses the correlation measurements or triggered observables; we leave such study to a future effort. The study along such direction may potentially provide more constraints on modeling initial states, thus helping our quantitative understanding of the transport properties of the hot and dense QGP produced in high energy heavy-ion collisions.

\section*{Acknowledgments}

We thank the Ohio State University group (Z. Qiu, C. Shen, H. Song and U. Heinz) for providing the corresponding initialization and hydrodynamic evolution codes. And we appreciate the support from the JET Collaboration. This work was supported by the U.S. Department of Energy Grant No. DE-FG02-05ER41367 and Natural Science Foundation of China under grant No. 11375072.

\section*{References}
\bibliography{SCrefs}

\providecommand{\newblock}{}
\begin{thebibliography}{10}
\expandafter\ifx\csname url\endcsname\relax
  \def\url#1{{\tt #1}}\fi
\expandafter\ifx\csname urlprefix\endcsname\relax\def\urlprefix{URL }\fi
\providecommand{\eprint}[2][]{\url{#2}}
% Bibliography created with iopart-num v2.1
% /biblio/bibtex/contrib/iopart-num

\bibitem{Kolb:2003dz}
Kolb P~F and Heinz U~W 2003  (\textit{Preprint} \eprint{nucl-th/0305084})

\bibitem{Adams:2003zg}
Adams J {\em et~al.\/} (STAR) 2004 {\em Phys. Rev. Lett.\/} {\bf 92} 062301
  (\textit{Preprint} \eprint{nucl-ex/0310029})

\bibitem{Aamodt:2010pa}
Aamodt K {\em et~al.\/} (The ALICE Collaboration) 2010 {\em Phys. Rev. Lett.\/}
  {\bf 105} 252302 (\textit{Preprint} \eprint{1011.3914})

\bibitem{ATLAS:2011ah}
Aad G {\em et~al.\/} (ATLAS Collaboration) 2012 {\em Phys. Lett.\/} {\bf B707}
  330--348 long author list - awaiting processing (\textit{Preprint}
  \eprint{1108.6018})

\bibitem{Luzum:2013yya}
Luzum M and Petersen H 2014 {\em J. Phys.\/} {\bf G41} 063102
  (\textit{Preprint} \eprint{1312.5503})

\bibitem{Alver:2010gr}
Alver B and Roland G 2010 {\em Phys. Rev.\/} {\bf C81} 054905
  (\textit{Preprint} \eprint{1003.0194})

\bibitem{Petersen:2010cw}
Petersen H, Qin G~Y, Bass S~A and Muller B 2010 {\em Phys. Rev.\/} {\bf C82}
  041901 (\textit{Preprint} \eprint{1008.0625})

\bibitem{Staig:2010pn}
Staig P and Shuryak E 2010  (\textit{Preprint} \eprint{1008.3139})

\bibitem{Qin:2010pf}
Qin G~Y, Petersen H, Bass S~A and Muller B 2010 {\em Phys. Rev.\/} {\bf C82}
  064903 (\textit{Preprint} \eprint{1009.1847})

\bibitem{Ma:2010dv}
Ma G~L and Wang X~N 2011 {\em Phys. Rev. Lett.\/} {\bf 106} 162301
  (\textit{Preprint} \eprint{1011.5249})

\bibitem{Xu:2010du}
Xu J and Ko C~M 2011 {\em Phys. Rev.\/} {\bf C83} 021903 (\textit{Preprint}
  \eprint{1011.3750})

\bibitem{Teaney:2010vd}
Teaney D and Yan L 2011 {\em Phys. Rev.\/} {\bf C83} 064904 (\textit{Preprint}
  \eprint{1010.1876})

\bibitem{Qiu:2011iv}
Qiu Z and Heinz U~W 2011 {\em Phys. Rev.\/} {\bf C84} 024911 (\textit{Preprint}
  \eprint{1104.0650})

\bibitem{Bhalerao:2011yg}
Bhalerao R~S, Luzum M and Ollitrault J~Y 2011 {\em Phys. Rev.\/} {\bf C84}
  034910 (\textit{Preprint} \eprint{1104.4740})

\bibitem{Floerchinger:2011qf}
Floerchinger S and Wiedemann U~A 2011 {\em JHEP\/} {\bf 1111} 100
  (\textit{Preprint} \eprint{1108.5535})

\bibitem{Adare:2011tg}
Adare A {\em et~al.\/} (PHENIX Collaboration) 2011 {\em Phys. Rev. Lett.\/}
  {\bf 107} 252301 (\textit{Preprint} \eprint{1105.3928})

\bibitem{Adamczyk:2013waa}
Adamczyk L {\em et~al.\/} (STAR Collaboration) 2013 {\em Phys. Rev.\/} {\bf
  C88} 014904 (\textit{Preprint} \eprint{1301.2187})

\bibitem{ATLAS:2012at}
Aad G {\em et~al.\/} (ATLAS Collaboration) 2012 {\em Phys. Rev.\/} {\bf C86}
  014907 (\textit{Preprint} \eprint{1203.3087})

\bibitem{Qin:2011uw}
Qin G~Y and Muller B 2012 {\em Phys. Rev.\/} {\bf C85} 061901
  (\textit{Preprint} \eprint{1109.5961})

\bibitem{Qiu:2012uy}
Qiu Z and Heinz U 2012 {\em Phys. Lett.\/} {\bf B717} 261--265
  (\textit{Preprint} \eprint{1208.1200})

\bibitem{Pang:2012uw}
Pang L, Wang Q and Wang X~N 2013 {\em Nucl. Phys.\/} {\bf A904-905} 811c--814c
  (\textit{Preprint} \eprint{1211.1570})

\bibitem{Gale:2012rq}
Gale C, Jeon S, Schenke B, Tribedy P and Venugopalan R 2013 {\em Phys. Rev.
  Lett.\/} {\bf 110} 012302 (\textit{Preprint} \eprint{1209.6330})

\bibitem{Luzum:2008cw}
Luzum M and Romatschke P 2008 {\em Phys. Rev.\/} {\bf C78} 034915
  (\textit{Preprint} \eprint{0804.4015})

\bibitem{Dusling:2007gi}
Dusling K and Teaney D 2008 {\em Phys. Rev.\/} {\bf C77} 034905
  (\textit{Preprint} \eprint{0710.5932})

\bibitem{Song:2010mg}
Song H, Bass S~A, Heinz U, Hirano T and Shen C 2011 {\em Phys. Rev. Lett.\/}
  {\bf 106} 192301 (\textit{Preprint} \eprint{1011.2783})

\bibitem{Pang:2012he}
Pang L, Wang Q and Wang X~N 2012 {\em Phys. Rev.\/} {\bf C86} 024911
  (\textit{Preprint} \eprint{1205.5019})

\bibitem{Andrade:2014swa}
Andrade R~P~G, Noronha J and Denicol G~S 2014  (\textit{Preprint}
  \eprint{1403.1789})

\bibitem{vanHees:2005wb}
van Hees H, Greco V and Rapp R 2006 {\em Phys. Rev.\/} {\bf C73} 034913
  (\textit{Preprint} \eprint{nucl-th/0508055})

\bibitem{Rapp:2009my}
Rapp R and van Hees H {\em published in R. C. Hwa, X.-N. Wang (Eds.), Quark
  Gluon Plasma 4 (World Scientific, 2010)\/}  111--206 (\textit{Preprint}
  \eprint{0903.1096})

\bibitem{Lin:2003jy}
Lin Z~W and Molnar D 2003 {\em Phys. Rev.\/} {\bf C68} 044901
  (\textit{Preprint} \eprint{nucl-th/0304045})

\bibitem{Greco:2003vf}
Greco V, Ko C~M and Rapp R 2004 {\em Phys. Lett.\/} {\bf B595} 202--208
  (\textit{Preprint} \eprint{nucl-th/0312100})

\bibitem{Oh:2009zj}
Oh Y, Ko C~M, Lee S~H and Yasui S 2009 {\em Phys. Rev.\/} {\bf C79} 044905
  (\textit{Preprint} \eprint{0901.1382})

\bibitem{He:2011qa}
He M, Fries R~J and Rapp R 2012 {\em Phys. Rev.\/} {\bf C86} 014903
  (\textit{Preprint} \eprint{1106.6006})

\bibitem{Cao:2013ita}
Cao S, Qin G~Y and Bass S~A 2013 {\em Phys. Rev.\/} {\bf C88} 044907
  (\textit{Preprint} \eprint{1308.0617})

\bibitem{Adare:2010de}
Adare A {\em et~al.\/} (PHENIX Collaboration) 2011 {\em Phys. Rev.\/} {\bf C84}
  044905 (\textit{Preprint} \eprint{1005.1627})

\bibitem{Tlusty:2012ix}
Tlusty D (STAR collaboration) 2013 {\em Nucl. Phys.\/} {\bf A904-905}
  639c--642c (\textit{Preprint} \eprint{1211.5995})

\bibitem{Grelli:2012yv}
Grelli A (ALICE Collaboration) 2013 {\em Nucl. Phys.\/} {\bf A904-905}
  635c--638c (\textit{Preprint} \eprint{1210.7332})

\bibitem{Caffarri:2012wz}
Caffarri D (ALICE Collaboration) 2013 {\em Nucl. Phys.\/} {\bf A904-905}
  643c--646c (\textit{Preprint} \eprint{1212.0786})

\bibitem{Uphoff:2012gb}
Uphoff J, Fochler O, Xu Z and Greiner C 2012 {\em Phys. Lett.\/} {\bf B717}
  430--435 (\textit{Preprint} \eprint{1205.4945})

\bibitem{Gossiaux:2010yx}
Gossiaux P, Aichelin J, Gousset T and Guiho V 2010 {\em J. Phys.\/} {\bf G37}
  094019 (\textit{Preprint} \eprint{1001.4166})

\bibitem{Nahrgang:2013saa}
Nahrgang M, Aichelin J, Gossiaux P~B and Werner K 2014 {\em Phys.Rev.\/} {\bf
  C90} 024907 (\textit{Preprint} \eprint{1305.3823})

\bibitem{Moore:2004tg}
Moore G~D and Teaney D 2005 {\em Phys. Rev.\/} {\bf C71} 064904
  (\textit{Preprint} \eprint{hep-ph/0412346})

\bibitem{Akamatsu:2008ge}
Akamatsu Y, Hatsuda T and Hirano T 2009 {\em Phys. Rev.\/} {\bf C79} 054907
  (\textit{Preprint} \eprint{0809.1499})

\bibitem{Young:2011ug}
Young C, Schenke B, Jeon S and Gale C 2012 {\em Phys. Rev.\/} {\bf C86} 034905
  (\textit{Preprint} \eprint{1111.0647})

\bibitem{Alberico:2011zy}
Alberico W, Beraudo A, De~Pace A, Molinari A, Monteno M {\em et~al.\/} 2011
  {\em Eur. Phys. J.\/} {\bf C71} 1666 (\textit{Preprint} \eprint{1101.6008})

\bibitem{Lang:2012cx}
Lang T, van Hees H, Steinheimer J and Bleicher M 2012  (\textit{Preprint}
  \eprint{1211.6912})

\bibitem{Cao:2011et}
Cao S and Bass S~A 2011 {\em Phys. Rev.\/} {\bf C84} 064902 (\textit{Preprint}
  \eprint{1108.5101})

\bibitem{Cao:2012jt}
Cao S, Qin G~Y and Bass S~A 2013 {\em J. Phys.\/} {\bf G40} 085103
  (\textit{Preprint} \eprint{1205.2396})

\bibitem{Rodriguez:2010di}
Rodriguez R, Fries R~J and Ramirez E 2010 {\em Phys. Lett.\/} {\bf B693}
  108--113 (\textit{Preprint} \eprint{1005.3567})

\bibitem{Renk:2011qi}
Renk T, Holopainen H, Auvinen J and Eskola K~J 2012 {\em Phys. Rev.\/} {\bf
  C85} 044915 (\textit{Preprint} \eprint{1105.2647})

\bibitem{Zhang:2012ik}
Zhang H, Song T and Ko C~M 2013 {\em Phys. Rev.\/} {\bf C87} 054902
  (\textit{Preprint} \eprint{1208.2980})

\bibitem{Guo:2000nz}
Guo X~F and Wang X~N 2000 {\em Phys. Rev. Lett.\/} {\bf 85} 3591--3594
  (\textit{Preprint} \eprint{hep-ph/0005044})

\bibitem{Majumder:2009ge}
Majumder A 2012 {\em Phys. Rev.\/} {\bf D85} 014023 (\textit{Preprint}
  \eprint{0912.2987})

\bibitem{Zhang:2003wk}
Zhang B~W, Wang E and Wang X~N 2004 {\em Phys. Rev. Lett.\/} {\bf 93} 072301
  (\textit{Preprint} \eprint{nucl-th/0309040})

\bibitem{Burke:2013yra}
Burke K~M {\em et~al.\/} (JET) 2014 {\em Phys. Rev.\/} {\bf C90} 014909
  (\textit{Preprint} \eprint{1312.5003})

\bibitem{Qin:2010mn}
Qin G~Y and Muller B 2011 {\em Phys.Rev.Lett.\/} {\bf 106} 162302
  (\textit{Preprint} \eprint{1012.5280})

\bibitem{Song:2007fn}
Song H and Heinz U~W 2008 {\em Phys. Lett.\/} {\bf B658} 279--283
  (\textit{Preprint} \eprint{0709.0742})

\bibitem{Song:2007ux}
Song H and Heinz U~W 2008 {\em Phys. Rev.\/} {\bf C77} 064901
  (\textit{Preprint} \eprint{0712.3715})

\bibitem{Qiu:2011hf}
Qiu Z, Shen C and Heinz U 2012 {\em Phys. Lett.\/} {\bf B707} 151--155
  (\textit{Preprint} \eprint{1110.3033})

\end{thebibliography}

\end{document}